\documentclass[aps,
prd,twocolumn,
prabib,
showpacs,nofootinbib,showkeys,10pt]{revtex4-2}
\usepackage[table]{xcolor}
\usepackage{graphicx} \usepackage{amsmath} \usepackage{amssymb}
\usepackage{amsfonts} \usepackage{bm}
\usepackage{array}
\usepackage{siunitx}

\usepackage[justification   = raggedright,singlelinecheck=false]{caption}

\usepackage{ytableau}
\usepackage{multirow}
\usepackage{mathtools}
\usepackage{tikz}

\begin{document}

\newcommand{\be}{\begin{equation}} \newcommand{\ee}{\end{equation}}
\newcommand{\bea}{\begin{eqnarray}}\newcommand{\eea}{\end{eqnarray}}

\title{Quantum walk search based   edge  detection of images} 

\author{Pulak Ranjan Giri} \email{pu-giri@kddi-research.jp}

\author{Rei Sato} \email{ei-satou@kddi.com}

\author{Kazuhiro Saito} \email{ku-saitou@kddi.com}

\affiliation{KDDI Research, Inc.,  Fujimino-shi, Saitama, Japan}

\begin{abstract} 
Quantum walk has emerged  as  an essential tool for searching  marked vertices on various graphs.   Recent advances  in the   discrete-time quantum walk  search algorithm  have enabled it  to effectively handle   multiple marked vertices,  expanding its range of  applications  further.   
In this article, we propose a novel application  of  this   advanced  quantum walk search algorithm  for  the  edge  detection of  images\textemdash   a critical task in digital image processing. 
Given  the probabilistic nature of  quantum computing,  obtaining  measurement  result    with a high success probability is essential alongside faster computation time. 
Our quantum walk search algorithm  demonstrates a high success probability   in detecting   the   image edges   compared to the existing  quantum edge detection methods and outperforms  classical edge detection methods with a quadratically faster  speed.  A small Qiskit  circuit  implementation of our method  using a one-dimensional quantum walk search  has  been executed in  Qiskit's    $qasm\_simulator$ and   $ibm\_sydney(fake)$ device.

 \end{abstract}


\date{\today}

\maketitle 



\section{Introduction} \label{in}
Edge detection is a crucial task  in  digital  image processing,  with  applications  across  various   fields  such as   medical imaging, forensic science,  material science,   and traffic surveillance.  Many  classical edge detection algorithms  exist \cite{sobel,pre,kir,can,niya},  but they struggle to keep up with the increasing size and volume of  image data.  These  methods are also  extremely  time-consuming,  as  they   need to examine  each pixel  individually  to determine the edges.   However, in some applications,  such as   self-driving vehicles \cite{thai},  real-time fast  processing of   edge detection is crucial for  a   safe driving.  Quantum computing \cite{ni} is considered  a promising  alternative   to efficiently  handle   large  image-datasets  faster than the classical methods.    

Quantum computing     utilizes    the   properties of  quantum mechanics, such as  entanglement, superposition,  and parallelism,    to speed  up       computational  processes.  Deutsch's  algorithm to distinguish  between  constant  and  balanced functions, Grover's  algorithm  \cite{grover1}  and its generalization \cite{giri} for  searching   marked elements  in  an unsorted database,   and  Shor's  algorithm \cite{shor1} for prime   factorization  are some of the most popular    quantum algorithms  known for their faster   computation   times   compared to their classical counterparts.

In the context of edge detection, several  quantum  algorithms, such as  QSobel \cite{zhang} and   Hadamard edge detection (HED), \cite{wei}  exist,  that  offer    exponential speedup   over   classical  methods. These quantum  edge detection (QED) algorithms   primarily  measure    pixel gradients  to determine the existence  of  edges.   However, they face challenges  when the intensity difference between neighboring  pixels is small, resulting in low probability amplitudes that  are   difficult to measure, especially in noisy conditions.   Since quantum computing  heavily relies  on  the  probability  for successful  outcomes, it  poses a significant challenge.

In this article,  we propose a new quantum edge detection method  to  address    this issue by using a  quantum walk search to amplify the  amplitudes of edge pixel intensities,  making them easier to detect and measure.  Our approach   involves  identifying   the edge pixels of a  digital image  by   searching  for the edges   as  a set of marked vertices.

Quantum walk search\textemdash  a generalization  of the Grover's  algorithm, has proven  significant  for  searching    on various graphs \cite{portugal}. 
In fact,   quantum walk (QW), the quantum analog  of  classical random walk, has  become  an essential tool  for     spatial search  on graphs.  
Both    the  continuous-time  and discrete-time quantum walk are  extensively employed  in the literature   \cite{portugal,childs,tulsi,amba3,wong1,giri3,tomo} to search for   marked vertices  on graphs faster than the   random walk.
Example   of graphs   on which spatial search have been successfully  implemented include:     $n$-dimensional square  lattice, complete graph,  $m$-partite complete graph, and $n$-dimensional  hypercube among  others.   Quantum walk exploits  quantum superposition to enhance search efficiency, making it a powerful tool for the spacial search on  various graph structures.

Many   applications  of quantum walk  leverage   their ability to search for the marked  vertices   faster than their   classical counterparts.   
For instance, a single marked vertex  on a two-dimensional grid with $N$ vertices can be  found    in  $\mathcal{O}(\sqrt{N \log N})$ time \cite{tulsi,amba3,wong1}.  Note that,  this time complexity is  slower by a factor of  $\sqrt{\log N}$   compared to the  optimal  speed of $\mathcal{O}(\sqrt{N})$ achieved by the  Grover search.  However, a quadratic speedup   of  $\mathcal{O}(\sqrt{N})$  can  be achieved in this  grid by adding  extra  long-range edges to the grid vertices \cite{giri3,tomo}.    

Searching for  the multiple marked  vertices,  specifically on a two-dimensional grid, is challenging  due  of the existence of {\it exceptional configurations} \cite{rivosh1,amba5,men}.  This   is a   set of specific arrangements  of marked vertices that   discrete-time  quantum walk search (DTQWS) algorithms with the standard coins cannot find.  Recent  research on lackadaisical quantum walk  \cite{giriepjd} shows that  modifying the Grover  coin operator to  flip  only the  phases of the  basis states of  marked vertices   with attached self-loops  can bypass this issue.  This modification allows DTQWS to successfully  search  for the   multiple marked vertices in  any configuration,  including the {\it exceptional configurations},    with very high success probability,   enabling  us to successfully exploit it for  the  quantum edge detection.  
\begin{figure}
 \includegraphics[width=0.45\textwidth]{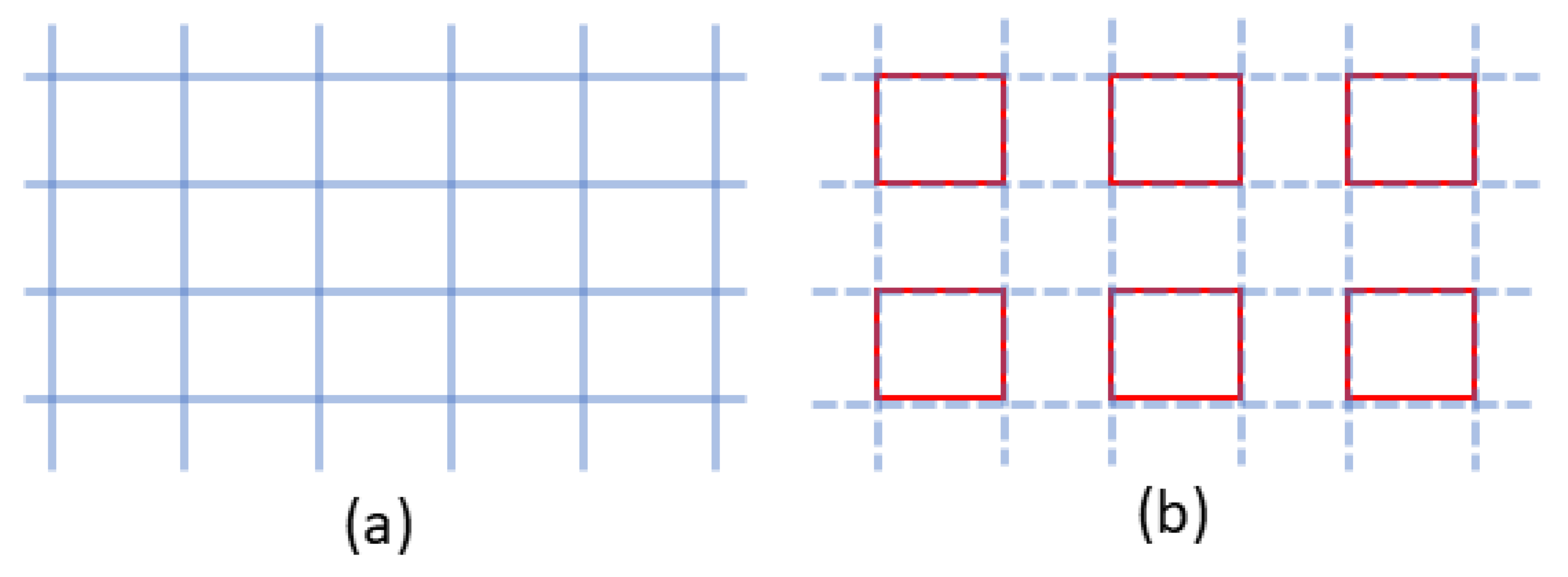}
          
       \caption{\bf{(a) A two-dimensional lattice of size $4\times 6$, (b) six $2\times 2$ blocks of the  $4\times 6$ two-dimensional lattice.}}
 \label{f1}
\end{figure}

This  article  is organized     as follows:   A  brief  discussion  on  the  quantum walk  search\textemdash  an essential  tool for our proposed  quantum edge detection method,   is  presented  in section    \ref{qw}.  In section   \ref{qc},  a quantum circuit  implementation   of this  detection method has  been performed in fake IBM quantum device. 
Results  of the  proposed edge detection method   are  analyzed   in section \ref{edge}. A  performance comparison of our edge detection method with other quantum edge detection methods   is done in section \ref{com}   and  finally  we conclude in section   \ref{con} with a discussion. 
\begin{figure}
 \includegraphics[width=0.45\textwidth]{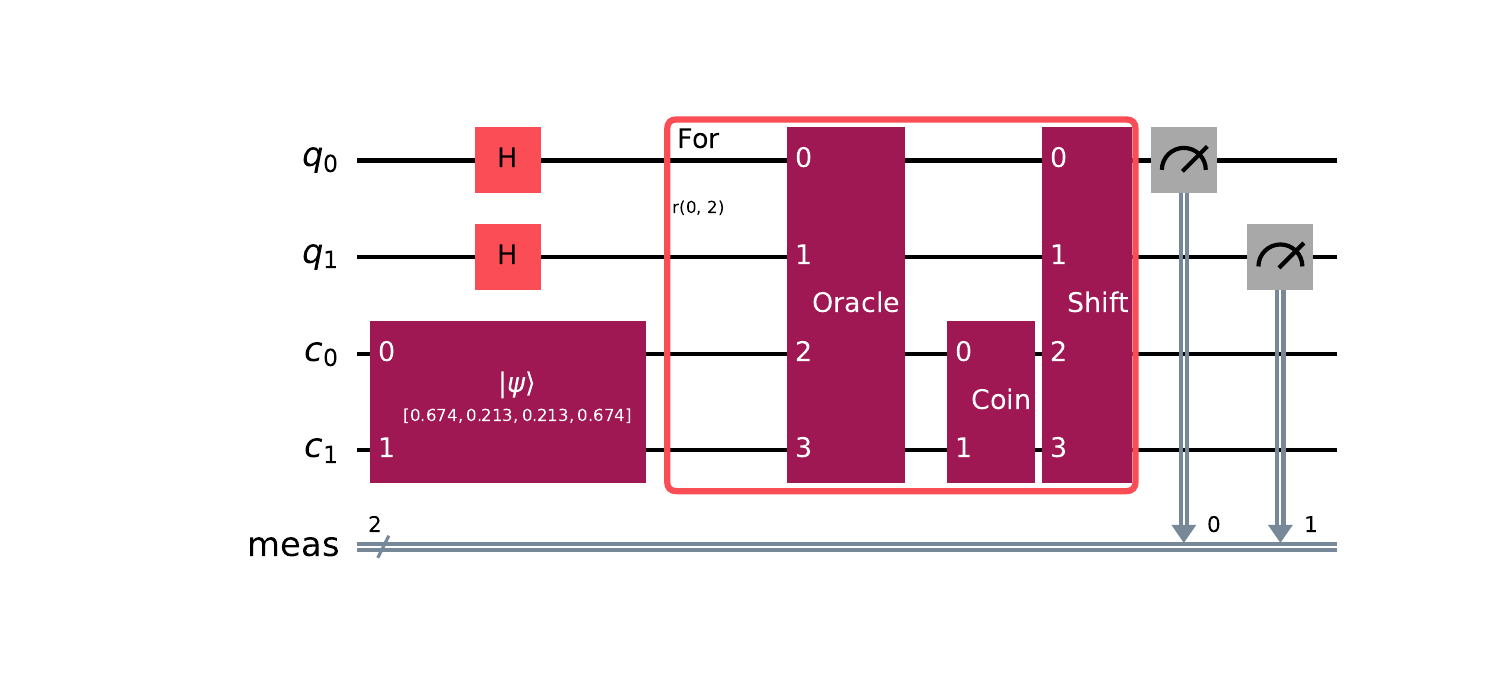}
          
       \caption{\bf{Qiskit circuit for quantum edge detection with quantum walk search on a cycle with four  vertices. Four dimensional coin  space, having four basis states:  left and right directions and  two self-loops with $\bf s= 0.1$, $\bf t=2$.}}
 \label{f2}
\end{figure}

\section{QW search on 2D grid} \label{qw}
 
Let us assume that   the pixel intensities $I(x,y)$  of a   $N_1 \times N_2$ digital image  are  located at the vertices of a  $N_1 \times N_2$ rectangular   lattice.  Fig. \ref{f1}(a) is an example of  a  $4\times 6$ two-dimensional lattices. 
Each vertex has  four regular  edges\textemdash right(r), left(l), up(u), and down(d). We also add one self-loop at each vertex,  so that  lackadaisical quantum walk can be implemented  on  the lattice.  Note that,  more than one self-loops \cite{wong1} can also be attached at each vertices for the lackadaisical quantum walk.  The vertices are the basis states of the $N= N_1 N_2$-dimensional Hilbert space,  $\mathcal{H}_V$,    and the edges form  the  $d= 5$-dimensional Hilbert space, $\mathcal{H}_C$,  of  the  coin.  Quantum walk starts  with an initial state,  
$|\psi_{in}\rangle = |\psi_{v}\rangle \otimes |\psi_{c}\rangle$, where 
\begin{eqnarray}
 |\psi_{v}\rangle =   \frac{1}{\sqrt{N}} 
\sum_{x=1}^{N_1} \sum_{y=1}^{N_2}   |x, y \rangle \,,
 \label{inv}
\end{eqnarray}
is the initial state in the vertex space with $(x, y)$ being the position of the vertices(pixels)  and 
\begin{eqnarray}
 |\psi_{c}\rangle   =   \frac{1}{\sqrt{4 +s}} 
\left( |r  \rangle  + |l  \rangle  + |u  \rangle  + |d  \rangle   + \sqrt{s} | s \rangle \right)\,,
 \label{in4} 
\end{eqnarray}
is the initial state in the coin space, where  $|r  \rangle,   |l  \rangle,   |u  \rangle,   |d  \rangle$ are the basis states for  the right, left, up and down  edges respectively.     $| s \rangle$ is the basis state  for the  directed self-loop  with weight $s$. 

The time evolution  of the initial state,  $|\psi_{in}\rangle$,  is governed by the unitary  operator,  $\mathcal{U} = S \mathcal{C}_G$,    composed  of  the  modified coin operator,  $\mathcal{C}_G$, followed by the  flip-flop shift  operator $S$.  
Let us assume that the  $M$ edge  pixels of an image, represented by   $M$  vertices,  belong  to  the  set  $\mathcal{T}_M$.  In the language of quantum walk   we have to then  find  $M$ marked vertices.  In this article, we use one-dimensional filter mask  \cite{geng} to determine the existence of edge pixels.   From the horizontal and vertical gradients $I_{h\pm}(x,y)$  and $I_{v\pm}(x,y)$:
\begin{eqnarray}\nonumber 
I_{h\pm}(x,y) &=& I(x,y) - I(x\pm1, y) \,,\\ \nonumber
I_{v\pm}(x,y) &=& I(x,y) - I(x, y\pm1) \,,
\label{grad}
\end{eqnarray}
we can evaluate the existence   of   edge  if  the gradient  $I_{max}(x,y)$ satisfies  the threshold condition   
\begin{eqnarray}
I_{max}(x,y) =  \mbox{max} \left( I_{h+}, I_{h-}, I_{v+}, I_{v-} \right) \geq a_{th}\,,
\label{thcon}
\end{eqnarray}
where $a_{th}$ is the pre-determined threshold value, that depends on the image.  Based on the result in  eq. (\ref{thcon}),  the marking of edges are done by the modified coin operator. 
\begin{figure}
  \centering
     \includegraphics[width=0.45\textwidth]{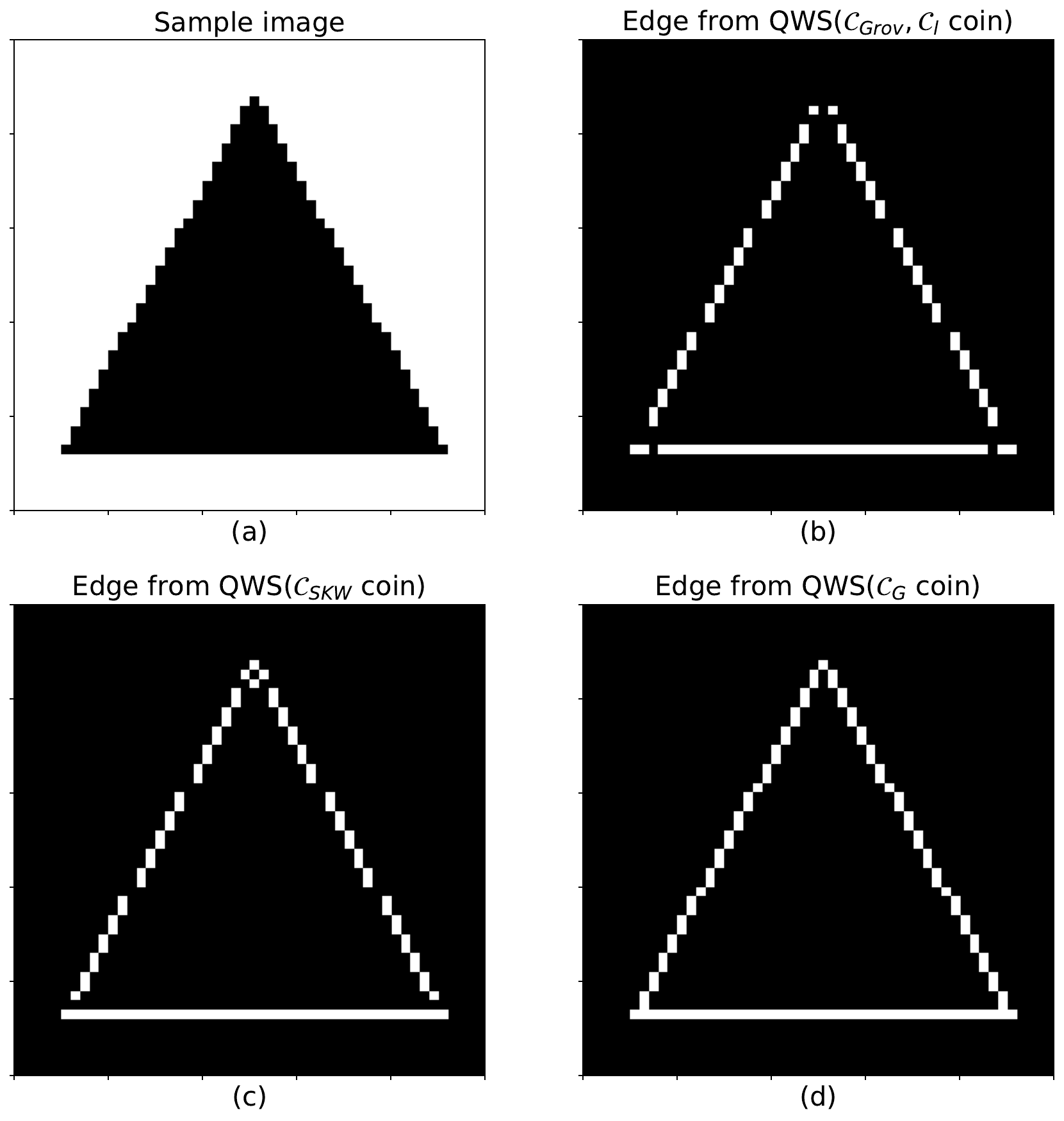}
          
       \caption{\bf {(a) $ \bf 50 \times 50$ sample image, and its   edge detection   from quantum walk search algorithm with  (b) regular or lackadaisical quantum walk with Grover coin  with $\bf s=0.01$ (c)  SKW coin and (d) $\bf \mathcal{C}_G$ coin with  $\bf s= 0.01$.  (b) -(d) are obtained by numerical analysis.}}
 \label{f3}
\end{figure}

In this article, we  are  using   the  following  modified coin   operator  \cite{giriepjd} 
\begin{equation}
 \mathcal{C}_G |x, y \rangle  \otimes | s \rangle  = 
 \begin{cases}
      ~~C|x, y \rangle  \otimes | b_c \rangle & \text{if  $(x,y)  \notin  \mathcal{T}_M$}\\
      ~~C|x, y \rangle  \otimes | b_c \rangle &\text{if  $(x,y)  \in  \mathcal{T}_M$; $ b_c \neq s$}\\
      - C| x, y \rangle  \otimes | b_c \rangle & \text{if  $(x,y)  \in  \mathcal{T}_M$;  $ b_c = s$}
\end{cases}  
\label{coing}           
\end{equation}
where  $C= 2|\psi_{v}\rangle \langle \psi_{v}| - \mathbb{I}_{d \times d}$  is the Grover  diffusion operator and $b_c$ is one of the basis states of the coin space.
Action of the shift operator on the basis states of the Hilbert space  $\mathcal{H} =  \mathcal{H}_V  \times \mathcal{H}_C$ is summarized as 
\begin{eqnarray} \nonumber 
S|x, y \rangle  \otimes | r \rangle  &=& |x+1, y \rangle  \otimes |l \rangle\,.\\ \nonumber
S|x, y \rangle  \otimes | l \rangle  &=& |x-1, y \rangle  \otimes |r \rangle\,,\\ \nonumber 
S|x, y \rangle  \otimes | u \rangle  &=& |x, y+1 \rangle  \otimes |d \rangle\,,\\  \nonumber
S|x, y \rangle  \otimes | d \rangle  &=& |x, y-1 \rangle  \otimes |u \rangle\,, \\
S|x, y \rangle  \otimes | s \rangle  &=&  |x, y \rangle  \otimes | s \rangle\,.
\label{shift}
\end{eqnarray}
The   success probability to find  all the $M$ marked  vertices(boundary pixels) of the  set $\mathcal{T}_M$,   after  $t$ time  steps,  is 
\begin{eqnarray} 
p_{s} = \sum_{ (x, y) \in \mathcal{T}_M} \sum_{ b_c} |\langle  x, y  | \otimes \langle b_c | \mathcal{U}^t |\psi_{in} \rangle|^2\,.
 \label{psucc}
 \end{eqnarray}
Measuring the output  will  give  edges of the image.  At this  present state of quantum computer,  it is hard to implement  the above discussed 2d DTQWS  method  in  NISQ device for the entire  image with  a reasonable-size.  However, in the next section we instead  implement a one-dimensional  DTQWS on $2\times 2$ blocks of an image, which requires   only four  qubits to implement our edge detection method.

\section{Quantum circuit implementation} \label{qc}
In this section we implement a Qiskit    circuit for the edge detection,  which is  suitable for  deployment in the current  IBM NISQ devices. 
Because of the limitations   of  the current   NISQ devices,  we design  a  small  quantum walk based  circuit  for our quantum search algorithm with $\mathcal{C}_G$ coin on the  one-dimensional periodic lattice, which can  be exploited  for the edge detection of a $N_1 \times N_2$ digital image.  
Let us assume  that  the length in both  directions   $N_1$ and $N_2$ are even. If the lengths are not even  for a digital image,  we can pad  appropriately to make them even.   Then,  we   divide the whole image  into $2 \times 2$  blocks, each having $4$ pixels.  For example,  an  $4\times 6$ two-dimensional lattice in fig. \ref{f1}(a) is divided into six $2\times 2$ red blocks as shown in fig. \ref{f1}(b). 
The $4$ pixel intensities of each block can be considered as the   vertices of a cycle  with length  N = 4.   We need  two qubits to represent three edges(left, right, self-loop).  However, one  extra basis state remains which we  assign  to another self-loop.  So, in this model  we have  two self-loops. 
Then the Hilbert space for the quantum state is   $\mathcal{H}_V  \times \mathcal{H}_C = \mathcal{C}^4 \times \mathcal{C}^4$.   Because of the various limitations of the current NISQ devices, it is important to keep the number of iterations  small, so that the  error related to  gate noise and decoherence  can be minimized.
For the  present study of the edge detection, we consider a Qiskit circuit with $t= 2$ iterations for  the quantum walk search, which is  shown  in fig.  \ref{f2}.  Once  the edges of  all the $2\times 2$ blocks are evaluated by the quantum circuit,  we  combine them to obtain the edge of the whole image.   Post-processing of the edge information is done with a threshold, $a_{th}$,  to obtain binary edge image.    The  one-dimensional QWS  for the $2\times 2$ blocks of the image  to obtain the edge is only done for the  $\bf qasm\_simulator$ and $\bf ibm\_sydney(fake)$ backend. Rest of the works are based of the two-dimensional  QWS of the entire image.

\section{Edge detection by DTQWS} \label{edge}
One of the important operations    in the evolution operator of the  DTQWS   is the  modified coin operator.  Several modified coins exist  in the literature, such as  the  Grover  coin, $\mathcal{C}_{Grov}$,  the coin for the lackadaisical quantum walk search, $\mathcal{C}_l$, SKW coin, $\mathcal{C}_{SKW}$,  and the  $\mathcal{C}_G$ coin. For  a  detailed discussion on these coins see ref. \cite{giriepjd}.   It is  known  that  there exist  {\it exceptional configurations}  in the search for  multiple  marked vertices, which  can not be found by   the   $\mathcal{C}_{Grov}$, $\mathcal{C}_{l}$ and $\mathcal{C}_{SKW}$ coins.  The success probabilities  of  the marked vertices remain  at their  initial values  in such configurations, despite repeated application of the evolution operator.  However,  $\mathcal{C}_{G}$, which has been described in section  \ref{qw},   can search any configurations with high success probability.  Moreover, it is the only coin which can efficiently search multiple vertices in one-dimensional periodic lattice, which is crucial for the  qiskit implementation in section \ref{qc}.  
A comparison among the results, obtained from the four coins   in fig. 2,  indicates  that  $\mathcal{C}_{G}$ is more suitable  for  edge detection. 
The  $50\times 50$ binary  image in fig. \ref{f3}(a)  has three  edge-images,  \ref{f3}(b) - 3(d),  obtained using  the four  coins.  We can see  $\mathcal{C}_{G}$  has  detected the  edge clearly in fig. \ref{f3}(d).  However, the edges  obtained by   $\mathcal{C}_{Grov}$ or  $\mathcal{C}_{l}$ coins in fig. \ref{f3}(b)  and by $\mathcal{C}_{SKW}$  coin  in fig. \ref{f3}(c) with  $2$ iterations have  several missing edge pixels. The edge detected by these  coins, except  $\mathcal{C}_{G}$,  with  iterations corresponding to highest success probability(not shown here) deteriorates even  further.  In this article,  post-processed edge images  are binarized  with  black corresponding  to $0$  and white corresponding  to $1$. 

\begin{figure}
  \centering
     \includegraphics[width=0.45\textwidth]{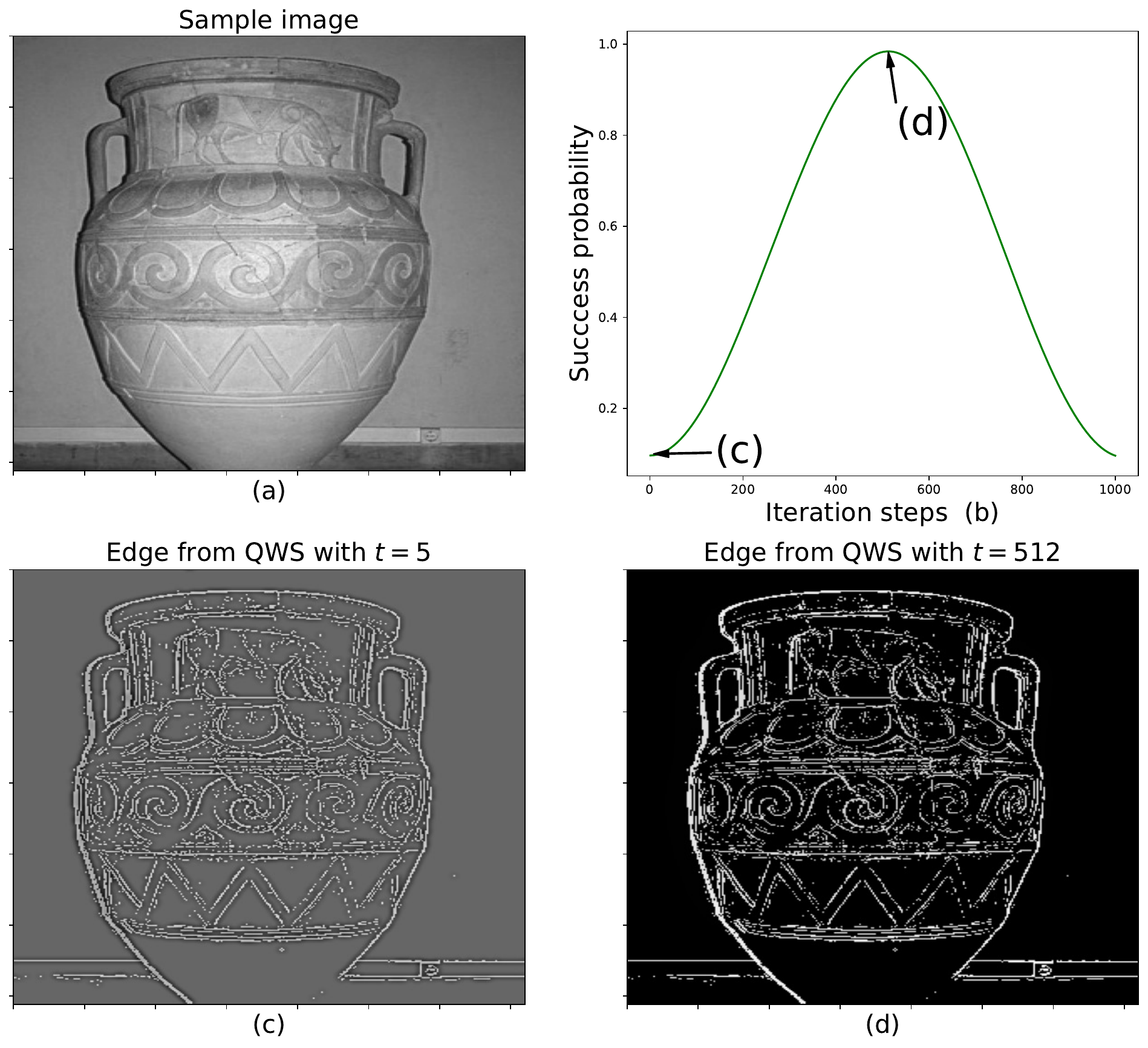}
          
       \caption{\bf{(a) $\bf 360 \times 306$ image  taken from the BSDS500 database  \cite{david},  (b) success probability as iteration steps for $\bf s=0.0001$,  edge detected image corresponding to (c) $\bf t=5$ and $\bf p_s \approx 0.1$ and  (d)  $\bf t=512$ and  $\bf p_s \approx 0.98$.  (b) -(d) are obtained by numerical analysis.}}
 \label{f4}
\end{figure}

Because $\mathcal{C}_{G}$  is more  suitable for  searching  multiple  marked vertices with high success probability, we use   this   coin for the  purpose of edge detection of an image.    Usually,  the  self-loop weight at each vertex of the graph needs to be fixed  at  an optimum value so that the success probability is maximized and time which corresponds  to highest success probability is considered for the edge detection. For example, a sample image  in fig. \ref{f4}(a) has  its edge obtained  after $t=5$ iterations in fig. \ref{f4}(c) and after $t=512$ iterations in fig. \ref{f4}(d).  As we can see from  the curve in fig. \ref{f4}(b) that  the edge in fig. \ref{f4}(d) corresponds to the maximum success probability, therefore having clear  edge compared to fig. \ref{f4}(c). 
Finally,  in fig. \ref{f5},  edge detection by  our quantum walk search technique has been obtained  by three different methods,  besides QSobel and Hadamard edge detection.  In  fig. \ref{f5}(d) numerical results from the two-dimensional  quantum walk search is shown. Then figs.  \ref{f5}(b) and  (c)  are obtained from the quantum circuit implementation  of the edge detection using one-dimensional quantum walk search  in section \ref{qc}.  The $qasm\_simulator$ result,  shown in fig. \ref{f5}(b),   is better than  the   result from  the noisy $ibm\_sydney$(fake) backend, in fig. \ref{f5}(c).  Figs. 5(e) and 5(f) are obtained with HED and QSobel method respectively. The graphs on the last row of fig. 5, obtained from the raw data of QWS, HED and  QSobel  methods respectively, show  that  our method  performs  better than  the other methods.  
\section{Performance comparison} \label{com}
In this section, performance of our  quantum edge detection model  over  other  significant   quantum edge detection models,  specifically  Hadamard edge detection and QSobel model,  is discussed.  Our model demonstrates  a markedly  high   total success probability, $p_s \approx 1$, outperforming  other models.     The  success probability for  $M$ edge vertices/targets   of  a digital image with  $N$ pixels can be optimized   by   tuning  the self-loop parameter $s$ accordingly.  Usually,  $s$   and $t$,  corresponding to the optimized $p_s$, depend on $N, M$, and  the dimension of the lattice.   A detailed numerical analysis  of  how the total success probability $p_s$  behaves as a function of the self-loop weight $s$  is studied in ref. \cite{giriepjd}.  Although  a  formal expression for the scaling of $p_s$ as a function of $N$ and $M$ is not known, the study   further shows that  even with a suitably chosen fixed $s$ the success probability $p_s$  remains reasonably  high  over a wide range of $N$ and $M$, as can be seen from figs. 3(b) and 5(b)  of ref. \cite{giriepjd}. Generally,  the lower the value of $s$, the  higher   the success probability $p_s$ and  time $t$, as can be seen from figs. 2(1d case) and fig. 4(2d case) of ref. \cite{giriepjd}. 

In the case of   Hadamard edge detection (HED) method,  
the encoding of  pixel intensities of a  $\sqrt{N} \times \sqrt{N}$ image  is done into  the   normalized   amplitudes of  a quantum state of the form 
\begin{eqnarray}
 |\psi_{in}\rangle =  
\sum_{i=0}^{2^{2n}-1}  c_i  | i \rangle \,,
 \label{hadin}
\end{eqnarray}
where amplitudes, $c_i$s,  are obtained by concatenating and normalizing   all the rows of  the  image's  pixel intensities.   This encoding method is referred to as     quantum probability image encoding \cite{gia}.   Here,  $2^{2n} = N$ represents  the  total number of pixels in the image.  
The process involves  applying a Hadamard gate to the first qubit  of the encoded state in   eq. (\ref{hadin}), which gives  
\begin{eqnarray}\nonumber 
 |\psi_{f}\rangle =  \frac{1}{\sqrt{2}} \sum_{i=0}^{2^{2n-1}-1}  \hspace{4.5cm} \\
\left(c_{2i} + c_{2i+1} \right) | 2i \rangle  +  \left(c_{2i} - c_{2i+1} \right) | 2i +1\rangle \,,
 \label{hadin1}
\end{eqnarray}
where only odd    basis states  encode  the edge information, while  all the even basis states are not useful for  edge detection  purposes.  An  edge is identified at position 
$(2i+1)$ if the gradient/amplitude   meets a threshold  condition, specifically  $\left(c_{2i} - c_{2i+1} \right)  \geq a_{th}$.  The gradients obtained in   eq. (\ref{hadin1})  are   $(c_0-c_1), (c_2-c_3), \cdots$.  The remaining  gradients/edges, such as  
$(c_1-c_2), (c_3-c_4), \cdots$,  can be obtained  by applying a shift operation  to  the initial state  in eq. (\ref{hadin}) and then continuing with the subsequent steps.  The total success probability for  detecting   edges  is  constrained by the limits 
\begin{eqnarray} \nonumber 
  p_{h} &=& \sum   | \frac{1}{\sqrt{2}} \left(c_{2i} - c_{2i+1} \right)|^2 \leq 1/2\\
  \tilde{p}_{h} &=& \sum  | \frac{1}{\sqrt{2}} \left(c_{2i+1} - c_{2i+2} \right)|^2  \leq 1/2 \,.
 \label{hadin2}
\end{eqnarray}
More than $1/2$ of the success probability is lost  with  the discarded even basis states. As a result, the total success   probability for the edge detection using   this  model is limited to   $p_h \leq 1/2$.  

\begin{figure}
  \centering
     \includegraphics[width=0.45\textwidth]{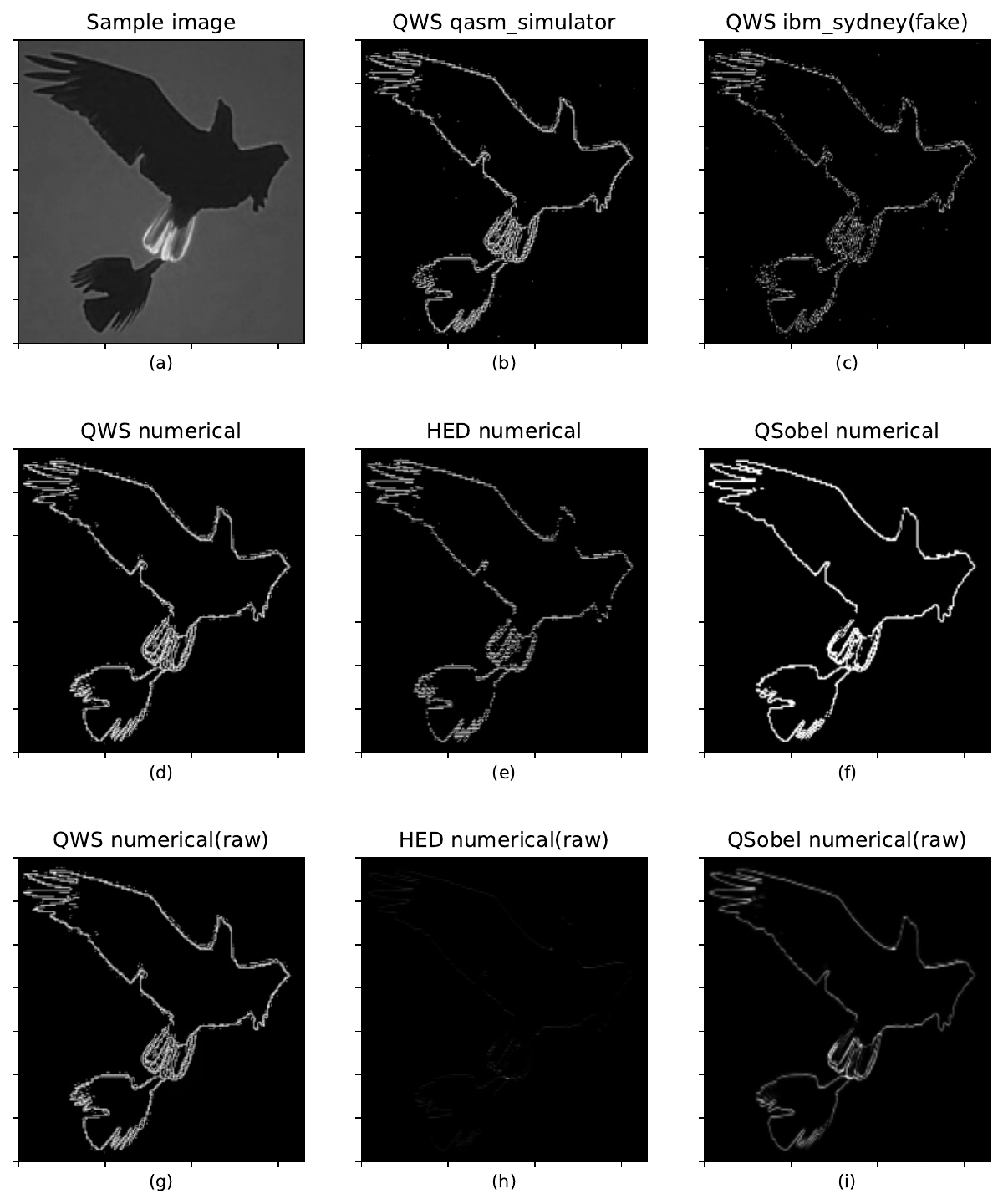}
          
       \caption{\bf{A sample (a) $\bf 330 \times 350$ image from  BSDS500 database  \cite{david} is used to obtain   its  edge   from  Qiskit circuit  for  the $2\times 2$ blocks of the image with (b) $\bf qasm\_simulator$(no noise),   $\bf  \bar{p}_s \approx 0.53$ and (c) $\bf ibm\_sydney$(fake) backend,   $\bf  \bar{p}_s \approx 0.40$,  $\bf s =0.1$, $\bf t=2$.
Also edge obtained with the  two-dimensional numerical QWS  of the   $\bf 330 \times 350$ image with  (d) QWS  with $\bf s= 0.0001$, $\bf t=791$,   $\bf  p_s \approx 0.98$, (e) HED,  $\bf  \bar{p}_h \approx 0.0031$, (f) QSobel,  $\bf  p_q \approx 0.0095$.
Edge obtained with raw the data(without post-processing) obtained from  numerical analysis of the entire image with  (g) QWS  with $\bf s= 0.0001$, $\bf t=791$,   $\bf  p_s \approx 0.98$, (h) HED,  $\bf  \bar{p}_h \approx 0.0031$, (i) QSobel,  $\bf  p_q \approx 0.0095$.}}
 \label{f5}
\end{figure}

In QSobel,  the initial state is  of the form \cite{zhang}
\begin{eqnarray}
 |\psi_{in}\rangle =  
\frac{1}{\sqrt{N}}\sum_{x=0}^{2^{n}-1}  \sum_{y=0}^{2^{n}-1} |0\rangle^{\otimes 9} |C_{xy}\rangle  | x,y \rangle \,,
 \label{qsobin}
\end{eqnarray}
where intensity of  image's  pixel is encoded  in the angle of the color qubit    $ |C_{xy}\rangle = \cos\theta_{xy} |0\rangle + \sin\theta_{xy} |1\rangle$. The final state, which contains the edge information,  is 
\begin{eqnarray}
 |\psi_{f}\rangle =  
\frac{1}{\sqrt{N}}\sum_{x=0}^{2^{n}-1}  \sum_{y=0}^{2^{n}-1}  |\Omega_{xy}\rangle  | x,y \rangle \,,
 \label{qsobin}
\end{eqnarray}
where qubit  $|\Omega_{xy}\rangle $  contains the color information  of the edge. 
The  total success probability to measure  $M$ edge pixels can be bounded   as 
\begin{eqnarray}
p_{q} =  M/N\,, 
 \label{qsobin1}
\end{eqnarray}
which is much less than the success probability of our edge detection model for $M << N$.  The actual success probability   $p_{q} =  \sum_{ (x,y)  \in \mathcal{T}_M} \sin^2 \theta_{xy}/N$  is even smaller than the upper bound $M/N$ obtained in eq.  (\ref{qsobin1}).

The effect of   higher success probability on  edge detection can be understood from the image edges   presented in the third  row of fig. \ref{f5}, which are obtained from the raw data(without post-processing). Numerical values of all the success probabilities are obtained from the raw data. Note that,  the success probability of our model  $p_s \approx 0.98$ is much higher  than   the success  probabilities  $\bar{p}_h = (p_h + \tilde{p}_h)/2  \approx 0.0031$ (HED), and $p_q \approx 0.0095$ (QSobel) respectively. Since our model has high success probability, the raw data plot in fig. 5(g) is almost same as its post-processed data plot in fig. 5(d). Whereas,  for  HED  and QSobel cases  edges  obtained from  the raw data plots in figs. 5(h) and  5(i) respectively are  very weak compared to their post-processed plots in figs. 5(e) and 5(f).  In  Qiskit circuit simulation,  the success probability is for the $2\times 2$ blocks only. So,  we have  obtained the average success probability over all the blocks present in the image.  Note that,  because of the noise  and other errors, the  average  success probability  $\bar{p}_s \approx 0.40$ for $\bf ibm\_sydney$(fake)  is  less than the  success probability  $\bar{p}_s \approx 0.53$ for the $\bf qasm\_simulator$(no noise).  We  iterated for $t=2$ in the Qiskit circuit because of the limitations of the simulator.  However, in the  pure numerical analysis of the one-dimensional  quantum walk search for the $2\times 2$ block,  a  much  higher success probability can be achieved.

\section{Conclusions} \label{con}
In this article,  we  proposed a  quantum walk search method  for   the  edge detection of digital  images.  Existing  quantum methods for    the edge detection often struggle with  low success probability, especially   in noisy environments.    Our approach,  which utilizes  quantum walk search,  boasts a higher  success probability, $p_s >> p_h, p_q$,  compared to other quantum edge detection models discussed in this article, making it more robust.   We have implemented a small quantum  circuit using Qiskit  that can run  on the  IBM quantum computer;  however,  a  full quantum advantage  requires  a quantum  circuit for  the entire   $N_1 \times N_2$ image in 2d,   which  current  NISQ quantum computers cannot handle.  Nonetheless,  because quantum walk search  is faster than  exhaustive search, our method  outperforms  classical edge detection methods in terms of speed. 

The presence of noise in the $\bf ibm\_sydney$(fake) simulator adversely affect the edge detection with reduced success probability  as can be seen from fig. 5(c).  In our study,  we have  not considered the presence of  noise in the sample images.  However,  the presence of noise in the sample images  should not  impact  the success probability of the image edge of our model assuming the oracle  can  mark  the target vertices/edges accurately.   Because, unlike other quantum edge detection models,  the initial state of our model $|\psi_{in}\rangle = |\psi_{v}\rangle \otimes |\psi_{c}\rangle$ is the equal superposition of all the vertex states, which is independent of   whether the sample image has noise or not.

For future work,   it would be worthwhile  to   explore  the impact  of error mitigations  on  the quantum walk search method for our edge detection.   Additionally, analyzing  the Qiskit  implementation of this  edge detection  method  using  quantum walk search  in $2$-dimensional  grid,  as   presented in  section \ref{qw},   and its deployment on  real quantum devices   would  be interesting.

\vspace{1cm}

Data availability Statement:  The data and database information  generated during and/or analyzed during the current study are included in the article.

\vspace{0.5cm}

Conflict of interest: The authors have no competing interests to declare that are relevant to the content of this article.


\end{document}